\documentclass[aoas,MSNbibl,nameyear,seceqn,dvips]{arximspdf}
\usepackage{dcolumn,multirow}
\usepackage{graphicx}

\doi{10.1214/12-AOAS589} 
\volume{7}
\issue{1}
\pubyear{2013}
\firstpage{391}
\lastpage{417}

\makeatletter
\newcolumntype{d}[1]{D{.}{.}{#1}}
\newcommand{\tnote}{\tabnoteref}
\newtheorem{lemma}{Lemma}
\newproclaim{remark}{Remark}

\makeatother

\begin{document}
\begin{frontmatter}

\title{Bootstrap inference for network construction with an application to a breast cancer microarray study\thanksref{T1}}
\runtitle{BINCO}
\thankstext{T1}{Supported by NIH Grants
R01GM082802 (SL, JP, PW), P01CA53996 (LH, PW), R01AG014358 (SL, LH),
P50CA138293 (LH, PW), U24CA086368 (PW), NSF Grant DBI-08-20854 (JP)
and DMS-10-07583 (JP).}

\begin{aug}
\author[A]{\fnms{Shuang} \snm{Li}\ead[label=e1]{Shuangli@fhcrc.org}},
\author[A]{\fnms{Li} \snm{Hsu}\ead[label=e2]{lih@fhcrc.org}},
\author[B]{\fnms{Jie} \snm{Peng}\ead[label=e3]{jiepeng@ucdavis.edu}}
\and
\author[A]{\fnms{Pei} \snm{Wang}\corref{}\ead[label=e4]{pwang@fhcrc.org}}
\runauthor{Li, Hsu, Peng and Wang}
\affiliation{Fred Hutchinson Cancer Research Center, Fred Hutchinson
Cancer Research Center, University of California, Davis
and Fred Hutchinson Cancer Research Center}
\address[A]{S. Li\\
L. Hsu\\
P. Wang\\
Fred Hutchinson Cancer Research Center\\
M2-B500, 1100 Fairview Ave N.\\
Seattle, Washington 98109\\
USA\\
\printead{e1}\\
\phantom{E-mail:\ }\printead*{e2}\\
\phantom{E-mail:\ }\printead*{e4}}
\address[B]{J. Peng\\
Department of Statistics\\
University of California, Davis\\
Mathematical Sciences Building\\
\quad One Shields Avenue\\
Davis, California 95616\\
USA\\
\printead{e3}}
\end{aug}

\received{\smonth{11} \syear{2011}}
\revised{\smonth{8} \syear{2012}}

%
\begin{abstract}
Gaussian Graphical Models (GGMs) have been used to construct genetic
regulatory networks where regularization techniques are widely used
since the network inference usually falls into a
high--dimension--low--sample--size scenario. Yet, finding the right
amount of regularization can be challenging, especially in an
unsupervised setting where traditional methods such as BIC or
cross-validation often do not work well. In this paper, we propose a
new method---Bootstrap Inference for Network COnstruction (BINCO)---to infer
networks by directly controlling the false discovery
rates (FDRs) of the selected edges. This method fits a mixture model
for the distribution of edge selection frequencies to estimate the
FDRs, where the selection frequencies are calculated via model
aggregation. This method is applicable to a wide range of
applications beyond network construction. When we applied our
proposed method to building a gene regulatory network with
microarray expression breast cancer data, we were able to identify
high-confidence edges and well-connected hub genes that could
potentially play important roles in understanding the underlying
biological processes of breast cancer.
\end{abstract}

%

\end{frontmatter}

\section{Introduction}\label{sec1}

The emergence of high-throughput technologies has\break made it
feasible to measure molecular signatures of thousands of
genes/\break proteins simultaneously. This provides scientists an
opportunity to study the global genetic regulatory networks,
shedding light on the functional interconnections among the
regulatory genes, and leading to a better understanding of
underlying biological processes. In this paper, we propose a network
building procedure for learning genetic regulatory networks. Our
work is motivated by an expression study of breast cancer (BC) that
aims to infer the network structure based on 414 BC tumor samples
[\citet{Loietal}]. The proposed method enables us to
detect high-confidence edges and well-connected hub genes that
include both those previously implicated in BC and novel ones that
may warrant further follow-up.

In practice, dependency structures of molecular activities such as
correlation matrix and partial correlation matrix have been used to
infer regulatory networks [\citet{Poletal02}, \citet{Kimetal06},
\citet{Varetal05},
Nie, Wu and Zhang (\citeyear{NieWuZha06})]. Such dependency structures are often
represented by graphical models in which nodes of a graph represent
biological components such as genes or proteins, and the edges
represent their interactions. These interactions may be indirect
(e.g., two genes are co-regulated by a third gene) or direct (e.g.,
one gene is regulated by another gene). For the latter case,
Gaussian Graphical Models (GGMs), which represent dependencies
between pairs of nodes conditioning on the remaining of nodes, are
often used.

For the data obtained from high-throughput technologies,
the number of nodes is typically much larger than the number of
samples, which is where the classical GGM theory [\citet{Whi90}]
generally fails [\citet{Fri89}, Scha\"{a}fer and
Strimmer (\citeyear{SchStr05})]. This large-$p$-small-$n$ scenario is usually
addressed by assuming that the conditional dependency structure is
sparse [\citet{Dobetal04}, \citet{LiGui06}, Meinshausen and Bu\"{u}hlmann
(\citeyear{MeiBuh06}), \citet{YuaLin07}, Friedman, Hastie and Tibshirani (\citeyear{FriHasTib08}),
\citet{Rotetal08},
Peng et al. (\citeyear{PenZhoZhu09})]. However, like many high-dimensional regularization problems,
finding the appropriate level of sparsity remains a challenge. This
is particularly true for network structure learning, since the
problem is unsupervised in nature. Traditional methods, such as
Bayes information criteria [\citet{Sch78}] and cross-validation, aim
to find a model that minimizes prediction error or maximizes a
targeted likelihood function. They tend to include many irrelevant
features [e.g., \citet{Efr04N2},
\citet{Efretal04},
Meinshausen and B\"{u}hlmann (\citeyear{MeiBuh06}) and
\citet{Penetal10}], and thus are not
appropriate for learning the interaction structures.

Choosing the amount of regularization by directly controlling the
false positive level would be ideal for structure learning.
Recently, a few model aggregation methods have been proposed, and
some of them provide certain control of false positives. For
example, \citet{Bach08} proposed
\textit{Bolasso}, which chooses variables that
are selected by all the lasso models [\citet{Tib96}] built on
bootstrapped data sets. In the context of network reconstruction,
\citet{Penetal10} proposed choosing
edges that are consistently selected across at least half of the
cross-validation folds. More recently, Meinshausen and
B\"{u}hlmann (\citeyear{MeiBuh10}) proposed the
\textit{stability selection}
procedure to choose
variables with selection frequencies exceeding a threshold. Under
suitable conditions, they derived an upper bound for the expected
number of false positives. In the same paper they also proposed the
randomized lasso penalty, which aggregates models from perturbing
the regularization parameters. Combined with stability selection,
randomized lasso achieves model selection consistency without
requiring the \textit{irrepresentable condition}
[\citet{ZhaYu06}] that is necessary for lasso to achieve model
selection consistency. In another work,
\citet{Wanetal11} proposed a modified
lasso regression---random lasso---by aggregating models based on
bootstrap samples and random subsets of variables. All these works
have greatly advanced research in model selection in the
high-dimensional regime. However, none of these methods provide
direct estimation and control of the false discovery rate (FDR).

In this paper, we address the problem of finding the right amount of
regularization in the context of high-dimension GGMs learning. In a
spirit similar to the aforementioned methods, we first obtain
selection frequencies from a collection of models built by
perturbing both the data and the regularization parameters. We then
model these selection frequencies by a mixture distribution to yield
an estimate of FDR on the selected edges, which is then used to
determine the cut-off threshold for the selection frequencies. This
framework is rather general, as it only depends on the empirical
distribution of the selection frequencies. Thus, it can be applied
to a wide range of problems beyond GGMs.

The rest of this paper is organized as follows. In Section~\ref{sec2} we
describe in detail the proposed method. In Section~\ref{sec3} an extensive
simulation study is conducted to compare the method with the
\textit{stability selection}
procedure and then evaluate its performance under different settings.
In Section~\ref{sec4} the method is illustrated by building a genetic
interaction network based on microarray expression data from a BC
study. The paper is concluded with some discussion in Section~\ref{sec5}.

\section{Method}\label{sec2}
\subsection{Gaussian graphical models}\label{sec2.1}

In a Gaussian Graphical Model (GGM) network construction is
defined by the conditional dependence relationships among the random
variables. Let $Y=(Y_1,\ldots,Y_p)$ denote a $p$-dimension random
vector following a multivariate normal distribution $N(0,\Sigma)$,
where $\Sigma$ is a $p\times p$ positive definite matrix. The
conditional dependence structure among $Y$ is represented by an
undirected graph $G=(U,E)$ with vertices $U=\{1,2,\ldots,p\}$
representing $Y_1,\ldots,Y_p$ and the edge set $E$ defined as
\[
E=\bigl\{(i,j)\dvtx Y_i \mbox{ and } Y_j \mbox{ are
dependent given }Y_{-\{i,j\}
}, 1\leq i,j\leq p\bigr\},
\]
where $Y_{-\{i,j\}}\equiv\{Y_k\dvtx k\not=i,j, 1\leq k\leq p\}$.
The goal of network construction is to identify the edge set $E$.
Under the normality assumption, the conditional independence between
$Y_i$ and $Y_j$ is equivalent to the partial correlation $\rho_{ij}$
between $Y_i$ and $Y_j$ given $Y_{-\{i,j\}}$ being zero. It is also
equivalent to the $(i,j)$ entry of the
\textit{concentration
matrix}
($\Sigma^{-1}$) being zero, that is,
$\sigma_{ij}\equiv (\Sigma^{-1})_{ij}=0$ [\citet{Dem72}, \citet{CoxWer96}],
since $\rho_{ij}= -\frac{\sigma_{ij}}{\sqrt{\sigma_{ii}\sigma_{jj}}}$.

There are two main types of approaches to fitting a GGM. One is the
maximum-likelihood-based approach, which estimates the concentration
matrix directly. The other is the regression-based approach, which
fits the GGM through identifying nonzero regression coefficients of
the following regression:
\[
Y_i=\sum_{j\not=i}\beta_{ij}Y_j+
\varepsilon_i,\qquad 1\leq i\leq p,
\]
where $\varepsilon_i$ is uncorrelated with $Y_{-i}=\{Y_k,k\not=i,1\leq
k \leq p \} $. The nonzero $\beta_{ij}$'s in the above regression
setting correspond to nonzero entries in the concentration matrix
since it can be shown that $\beta_{ij}=-\sigma_{ij}/\sigma_{ii}=\rho_{ij}\sqrt{\sigma_{jj}/\sigma_{ii}}$. In both
approaches, there are $O(p^2)$ parameters to estimate, which
requires proper regularization on the model if $p$ is larger than
the sample size $n$. This can be achieved by making a
\textit{sparsity assumption}
on the network
structure, that is, assuming that most pairs of variables are
conditionally independent given all other variables. Such an
assumption is reasonable for many real life networks, including
genetic regulatory networks [\citet{Garetal03},
Jeong et al. (\citeyear{Jeoetal11}),
\citet{Tegetal}]. Methods have been developed along these lines by using $L_1$
regularization. For example, \citet{YuaLin07} proposed a sparse
estimator of the concentration matrix via maximizing the $L_1$
penalized log-likelihood. Efficient algorithms were subsequently
developed to fit this model with high-dimensional data [Friedman, Hastie and Tibshirani
(\citeyear{FriHasTib08}),
\citet{Rotetal08}]. For regression-based approaches,
Meinshausen and B\"{u}hlmann (\citeyear{MeiBuh06}) considered the
neighborhood selection estimator by minimizing $p$ individual loss
functions
%
\begin{equation}\label{eq2.1}
L^{(i)}(\beta,Y)=\frac{1}{2}\biggl\Vert Y_i-\sum
_{j:j\not=i}\beta_{ij}Y_j\biggr\Vert^2+
\lambda\sum_{j:j\not=i}|\beta_{ij}|,\qquad i=1,\ldots,p,
\end{equation}
while Peng et al. (\citeyear{PenZhoZhu09}) proposed the
\textit{space}
algorithm by minimizing the joint loss
%
\begin{equation}\label{eq2.2}
L(Y,\theta)=\frac{1}{2}\Biggl\{\sum_{i=1}^p
\biggl\Vert Y_i-\sum_{j:j\not
=i}\sqrt{
\frac{\sigma_{jj}}{\sigma_{ii}}}\rho_{ij} Y_j\biggr\Vert^2\Biggr\}
+\lambda\sum_{1\leq i<j\leq p}|\rho_{ij}|.
\end{equation}

From objective functions (\ref{eq2.1}) and (\ref{eq2.2}), it is clear that the
selected edge set depends on the regularization parameter $\lambda$.
Since the goal here is to recover the true edge set, ideally
$\lambda$ should be determined based on considerations such as FDR
and power with respect to edge selection. Moreover, when the sample
size is limited, a model-aggregation-based strategy can improve the
selection result compared to simply tuning the regularization
parameter. Thus, in the following section, we introduce a new
model-aggregation-based procedure that selects edges based on
directly controlling the FDRs.

Throughout the rest of this paper, we refer to the set of all pairs
of variables as the
\textit{candidate edge
set}
(denoted by $\Omega$), the subset of
those edges in the true model as the
\textit{true edge
set}
(denoted by $E$) and the rest as the
\textit{null edge set}
(denoted by~$E^c$). We denote the
size of a set of edges by $|\cdot|$. Note that $\Omega=E\cup E^c$
and the total number of edges in $\Omega$ is $N_{\Omega
}=|\Omega|=p(p-1)/2$.

\subsection{Model aggregation}\label{sec2.2}

Consider a good network construction procedure, where good
is in the sense that the true edges are stochastically more likely
to be selected than the null edges. Then it would be reasonable to
choose edges with high selection probabilities. In practice, these
selection probabilities can be estimated by the selection
frequencies over networks constructed based on perturbed data sets.
In the following, we formalize this idea.

Let $A(\lambda)$ be an edge selection procedure with a
regularization parameter $\lambda$ and $S^{\lambda}(Y)\equiv
S^{\lambda}(A(\lambda),Y)$ be the set of selected edges by
applying $A(\lambda)$ to data $Y$. The
\textit{selection
probability}
of edge $(i,j)$ is defined as
\[
p_{ij}= E \bigl(I\bigl\{(i,j)\in S^{\lambda}(Y)\bigr\} \bigr),
\]
where
$I\{\cdot\}$ is the indicator function. Let $R(Y)$ be the space of
resamples from $Y$ (e.g., through bootstrapping or subsampling). For
a random resample $Y'$ from $R(Y)$, we define
\[
\tilde{p}_{ij}= E \bigl(I\bigl\{(i,j)\in S^{\lambda}
\bigl(Y'\bigr)\bigr\} \bigr)= E \bigl(E \bigl(I\bigl\{(i,j)\in
S^{\lambda}\bigl(Y'\bigr)\bigr\}\mid Y \bigr) \bigr).
\]
In many cases (see Section C in the supplemental article [Li et al. (\citeyear{Lietal})]), $p_{ij}$'s and $\tilde{p}_{ij}$'s are close. For these
cases, we can estimate $p_{ij}$ by the
\textit{selection
frequency}
$X_{ij}$, which is the
proportion of $B$ resamples in which the edge $(i,j)$ is selected:
%
\begin{eqnarray}\label{eq2.3}
X_{ij}^{\lambda}&\equiv& X_{ij} \bigl(A(
\lambda);Y^1,\ldots,Y^B \in R (Y) \bigr)
\nonumber
\\[-8pt]
\\[-8pt]
\nonumber
&=&
\frac{1}{B}\sum_{k=1}^BI \bigl\{(i,j)
\in S^{\lambda
}\bigl(Y^k\bigr) \bigr\},\qquad  1\leq i<j\leq p.
\end{eqnarray}

The aggregation-based procedures for choosing edges of large
selection frequencies can be represented as
\[
S_c^{\lambda}= \bigl\{(i,j)\dvtx X_{ij}^{\lambda}
\geq c\bigr\}\qquad \mbox{for } c\in(0,1].
\]
$S_c^{\lambda}$ is reasonable as long as most true
edges have selection frequencies greater than or equal to $c$ and
most null edges have selection frequencies less than~$c$. Ideally, we
want to find a threshold $c$ satisfying
%
\begin{equation}\label{eq2.4}\qquad
\operatorname{Pr} \biggl( \biggl\{\bigcap_{(i,j)\in E} \bigl
\{X_{ij}^{\lambda}\geq c\bigr\} \biggr\} \cap \biggl\{\bigcap
_{(i,j)\in E^c}\bigl\{X_{ij}^{\lambda}< c\bigr\}
\biggr\} \biggr)\rightarrow1\qquad \mbox{as }n\rightarrow\infty,
\end{equation}
so that the corresponding procedure $S_c^{\lambda}$ is consistent,
that is, $\operatorname{Pr}(S_c^{\lambda} =E)\rightarrow1$. In fact, if $A(\lambda)$
is selection consistent and $p_{ij}-\tilde{p}_{ij}\rightarrow0$, then
%
\begin{equation}\label{eq2.5}\qquad
\operatorname{Pr} \biggl( \biggl\{ \bigcap_{(i,j)\in E}\bigl
\{X_{ij}^{\lambda}=1\bigr\} \biggr\} \cap \biggl\{ \bigcap
_{(i,j)\in E^c}\bigl\{X_{ij}^{\lambda}=0\bigr\} \biggr\}
\biggr)\rightarrow1\qquad \mbox{as }n\rightarrow\infty,
\end{equation}
and thus any $c \in(0,1]$ satisfies (\ref{eq2.4}). Note that (\ref{eq2.4}) is in
general a much weaker condition than (\ref{eq2.5}), which suggests that we
might find a consistent $S_c^{\lambda}$ even when $A(\lambda)$ is not
consistent.

\begin{figure}
\centering
\begin{tabular}{@{}cc@{}}

\includegraphics{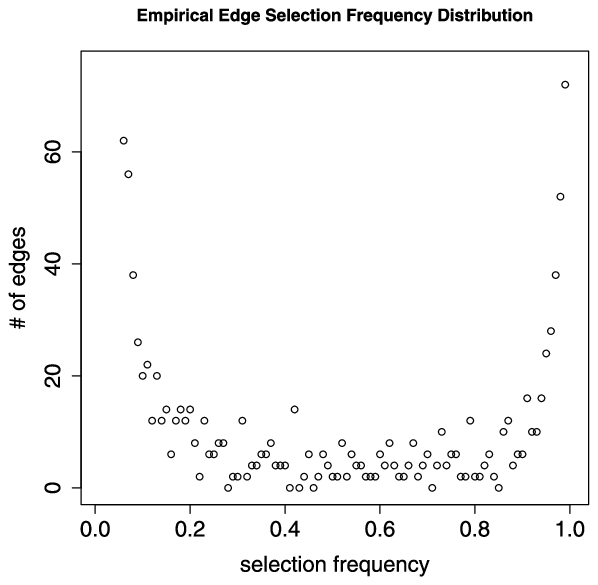}
 & \includegraphics{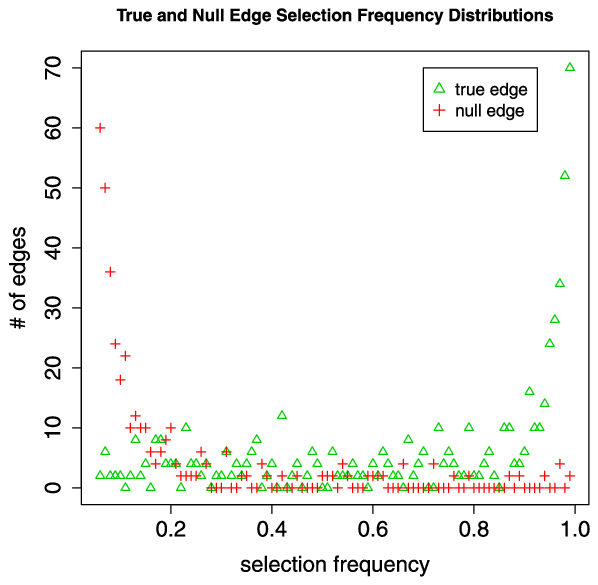}\\
\footnotesize{(a)} & \footnotesize{(b)}
\end{tabular}
\caption{The distributions of selection frequencies based on a
simulated data set. \textup{(a)} The distribution of selection frequencies of
all edges. \textup{(b)} Distributions of selection frequencies of null and
true edges, respectively (note that these are not observable in
practice). Simulation is based on a power-law network with $p=500$,
$n=200$, and the number of true edges is 495. The space
algorithm with $\lambda=135$ is
used as the original nonaggregation procedure $A(\lambda)$. For
illustrating the tail behavior of these distributions more
effectively, we only show them on the selection frequency range
[0.06, 1], as there are too many edges with selection frequency less
than 0.06.}\label{fig1}
\end{figure}

\begin{figure}

\includegraphics{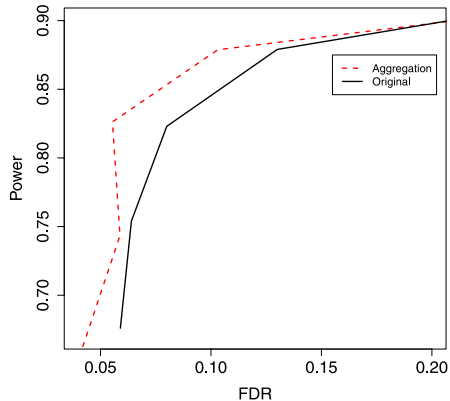}

\caption{Power and FDR of the aggregation-based procedure
$S_c^{\lambda}$ with cutoff $c=0.5$ and the original procedure
$A(\lambda)$ for $\lambda= 96, 114, 135, 160$, with the rest of
settings the same as in Figure~\protect\ref{fig1}.} \label{fig2}
\end{figure}

For the finite data case, an aggregation-based procedure could also
perform better than the original procedure, as illustrated by the
following simulation example (the simulation setup is provided in
Section~\ref{sec3}). Figure~\ref{fig1}(a) shows the empirical distribution of
selection frequencies based on a simulated data set and Figure~\ref{fig1}(b)
shows the empirical distributions of true edges (green triangles)
and null edges (red crosses). Note that most null edges have low
selection frequencies $<0.4$, while most true edges have large
selection frequencies $>0.6$. This suggests that with a properly
chosen $c$ (say, $c\in[0.4,0.6]$), $S_c^{\lambda}$ will select
mostly true edges and only a small number of null edges. In fact, by
simply choosing the cutoff $c=0.5$, $S_c^{\lambda}$ outperforms
$A(\lambda)$ in both FDR and power (Figure~\ref{fig2}).

\subsection{Modeling selection frequency}\label{sec2.3}

Now we introduce a mixture model, similar in spirit to \citet{Efr04N1},
for estimating the FDR of an aggregation-based procedure
$S_c^{\lambda}$. We will use this estimate to choose the optimal
$c$ and $\lambda$ by controlling FDR while maximizing power. Assume
that the selection frequencies $\{X_{ij}^{\lambda},(i,j)\in\Omega
\}$,\vspace*{1pt} generated from $B$ resamples, fall into two categories, ``true''
or ``null,'' depending on whether $(i,j)$ is a true edge or a null
edge. Let $\pi$ be the proportion of the true edges. We also assume
that $X_{ij}^{\lambda}$ has density $f_1^{\lambda}(x)$ or
$f_0^{\lambda}(x)$ if it belongs to the ``true'' or the ``null''
categories, respectively. Note that both $f_1^{\lambda}$ and
$f_0^{\lambda}$ depend on the sample size $n$, but such dependence
is not explicitly expressed in order to keep the notation simple.
The mixture density for $X_{ij}^{\lambda}$ can be written as
%
\begin{equation}\label{eq2.6}
f^{\lambda}(x)=(1-\pi)f_0^{\lambda}(x)+\pi
f_1^{\lambda}(x), \qquad x\in \{0,1/B,2/B,\ldots,1\}.
\end{equation}
Based on this mixture model, the (positive) FDR [\citet{Sto03}] of the
aggregation-based procedure $S_c^{\lambda}$ is
%
\begin{equation}\label{eq2.7}
\operatorname{FDR}\bigl(S_c^{\lambda}\bigr)=\operatorname{Pr} \bigl((i,j)\in
E^c|(i,j)\in S_c^{\lambda} \bigr)=\frac{\sum_{x\geq c}(1-\pi)f_0^{\lambda}(x)}{\sum_{x\geq
c}f^{\lambda}(x)}.
\end{equation}
Given an estimate $\widehat{\operatorname{FDR}}(S_c^{\lambda})$ (which will be
discussed below) from (\ref{eq2.7}), the number of true edges in $S_c^{\lambda
}$ can be estimated by
%
\begin{equation}\label{eq2.8}
\widehat{N}_{E}\bigl(S_c^{\lambda}
\bigr)=\bigl|S_c^{\lambda}\bigr| \bigl( 1-\widehat {\operatorname{FDR}}
\bigl(S_c^{\lambda}\bigr) \bigr),
\end{equation}
which can be used to compare the power of $S_c^{\lambda}$ across
various choices of $c$ and $\lambda$, as the total number of true edges
is a constant. Consequently, for a given targeted FDR level $\alpha$,
we first seek for the optimal threshold $c$ for each $\lambda\in
\Lambda $,
%
\begin{equation}\label{eq2.9}
c^*(\lambda)=\operatorname{min}\bigl\{c\dvtx \widehat{\operatorname{FDR}}\bigl(S_c^{\lambda}
\bigr)\leq\alpha\bigr\},
\end{equation}
and then we find the optimal regularization parameter
%
\begin{equation}\label{eq2.10}
\lambda^* = \mathop{\operatorname{argmax}}_{\lambda\in\Lambda} \widehat {N}_{E}
\bigl(S_{c^*(\lambda)}^{\lambda}\bigr),
\end{equation}
such that the corresponding procedure $S_{c^*(\lambda^*)}^{\lambda^*}$
achieves the largest power among all competitors with estimated FDR not
exceeding $\alpha$.

The above procedure depends on a good FDR estimate, which in turn
requires good estimates of the mixture density $f^{\lambda}$ and
its null-edge contribution $(1-\pi)f_0^{\lambda}$. A natural
estimator of $f^{\lambda}$ is simply the empirical selection
frequencies, that is,
\[
\hat{f}^{\lambda} \biggl(\frac{k}{B} \biggr)=\frac{n^{\lambda
}_k}{N_{\Omega}},\qquad
k=0,1,\ldots,B,
\]
where $N_{\Omega}=p(p-1)/2$ is the total number of candidate edges and
$n_k^{\lambda}=|\{(i,j)\dvtx X_{ij}^{\lambda}=k/B\}|$ is the number of
edges with selection frequencies equal to $k/B$.

Before describing an approach to estimating $\pi$ and $f_0^{\lambda
}$, we note two observations from Figure~\ref{fig1}(b). First, the
contribution from the true edges to the mixture density $f^{\lambda}$ is small in the
range where the selection frequencies are small. Second, the empirical
distribution of $f_0^{\lambda}$ is monotonically decreasing. These can
be formally summarized as the following condition.

\begin{pc*}
There exist $V_1$ and $V_2$, $0<V_1<V_2<1$, such that as
$n\rightarrow\infty$:
\begin{longlist}[(C1)]
\item[(C1)] $ f_1^{\lambda}\rightarrow0$ on $(V_1,V_2]$;

\item[(C2)] $f_0^{\lambda}$ is monotonically decreasing on $(V_1,1]$.
\end{longlist}
\end{pc*}

This
\textit{proper condition}
is satisfied by a class of procedures as described in the lemma below
(the proof is provided in the \hyperref[app]{Appendix}).

\begin{lemma}\label{le1}
A selection procedure satisfies the
\textit{proper condition}
if, as the sample size increases, $\tilde{p}_{ij}$ tends to one
uniformly for all true edges and has a limit superior strictly less
than one for all null edges.
\end{lemma}

\begin{remark}\label{rem1} It is easy to verify that all consistent procedures
applied to subsampling resamples satisfy the condition in Lemma~\ref{le1}.
Other examples are procedures that use randomized lasso penalties
[Meinshausen and B\"{u}hlmann (\citeyear{MeiBuh10})]. See Section~\ref{sec2.5} for
more details.
\end{remark}

The
\textit{proper condition}
motivates us to
estimate $\pi$ and $f_0^{\lambda}$ by fitting a parametric model
$g_{\theta}$ for $f^{\lambda}$ in the region $(V_1, V_2]$ and then
extrapolating the fit to the region $(V_2,1]$. This is because if C1
is satisfied, then $(1-\pi)f_0^{\lambda}$ can be well approximated
based on the empirical mixture density from the region $(V_1 ,
V_2]$. If C2 is also satisfied, the extrapolation of $g_{\theta}$
will be a good approximation to $(1-\pi)f_0^{\lambda}$ on
$(V_2,1]$ for a reasonably chosen family of $g_{\theta}$.

We choose the parametric family as follows. Given $\tilde{p}_{ij}$,
it is natural to model the selection frequency by a (rescaled)
binomial distribution, denoted by $b_1(\cdot|\tilde{p}_{ij})$, due
to the independent and identical nature of resampling conditional on
the original data. Moreover, we use a
\textit{powered beta}
distribution [i.e., the
distribution of $Q^\gamma$ where $Q\sim$ beta$(a,b)$, $a,b,r>0$]
as the prior for $\tilde{p}_{ij}$'s, denoted by $b_2(\cdot|\theta
)$ with $\theta=(a,b,r)$. This is motivated by the fact that the
beta family is a commonly used conjugate prior for the binomial
family, and the additional power parameter~$\gamma$ simply provides
more flexibility in fitting. Thus, the distribution of selection
frequencies of null edges is modeled as
\[
h_{\theta}(x) =\int_0^1
b_1(x|\tau)b_2(\tau|\theta)\,d \tau.
\]
The null-edge contribution $(1-\pi)f_0^{\lambda}$ can be estimated by
fitting $h_{\theta}$ to the empirical mixture density $\hat{f}^{\lambda
}$ in the
\textit{fitting range}
$(V_1,V_2]$, which, in practice, is determined based on the shape of
$\hat{f}^{\lambda}$ (details are given in Section~\ref{sec2.4}). Specifically,
we estimate $\pi$ and $f_0^{\lambda}$ by $\hat{\pi}$ and $h_{\hat
{\theta}}$, via
%
\begin{equation}\label{eq2.11}
(\hat{\pi},\hat{\theta})=\mathop{\operatorname{argmin}}_{\pi, \theta}L \bigl(\hat
{f}^{\lambda}(\cdot),(1-\pi)h_{\theta}(\cdot) \bigr),
\end{equation}
where $L(f,g)\equiv-\sum_{x\in(V_1,V_2]}[f(x)\operatorname{ log }g(x)]$,
which amounts to the Kullback--Leibler distance.

\subsection{Proper regularization range}\label{sec2.4}

\begin{figure}

\includegraphics{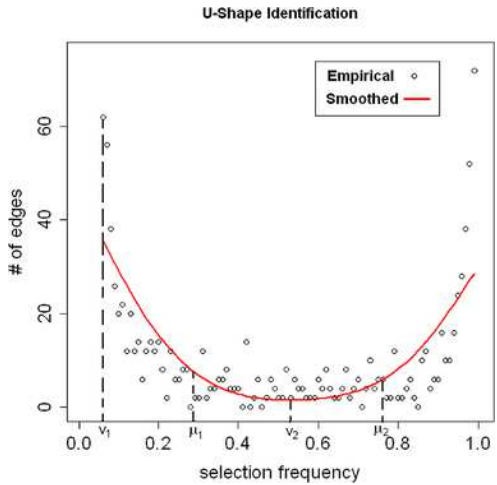}

\caption{An illustration for the proposed U-shape identification
procedure. The empirical distribution ($\hat{f}^{\lambda}$) is the
same as the one in Figure~\protect\ref{fig1}. The smooth curve ($\tilde{f}^{\lambda}$)
is fitted by the R-function $\tt{smooth.spline}$ with df${}=4$.
Locations of $v_1$, $v_2$, $\mu_1$ and $\mu_2$ are found by following
steps in the U-shape detection procedure.} \label{fig3}
\end{figure}

Following what we propose in Section~\ref{sec2.3}, we can evaluate
the aggregation-based procedure $S_{c}^{\lambda}$ for different
choices of $(\lambda,c)$ with regard to model--selection--based
criteria: the FDR and the number of selected true edges. For the
range of $\lambda$, we consider those that yield
``\textit{U-shaped}''
empirical distributions of
selection frequencies, that is, $\hat{f}^{\lambda}$ decreases in
the small-selection-frequency range and then increases in the
large-selection-frequency range [see Figure~\ref{fig1}(a) and Figure~\ref{fig3} for
examples of ``U-shaped'' distribution]. The decreasing trend is
needed for the
\textit{proper condition}
to hold,
while the increasing trend helps to control the FDR, since an
$S_{c}^{\lambda}$ with $\operatorname{FDR}\leq\alpha$ implies, by (\ref{eq2.7}), that
%
\begin{equation}\label{eq2.12}
\sum_{x\geq c}f^{\lambda}(x)\geq\frac{(1-\pi)\sum_{x\geq
c}f_0^{\lambda}(x)}{\alpha}.
\end{equation}

\textit{\hspace*{-12pt}\begin{tabular}{@{}l p{340pt}@{}}
\hline
\multicolumn{2}{c}{\begin{bfseries}\textbf{U-shape detection procedure}\end{bfseries}} \\
\hline
1. & \textbf{INPUT} $\hat{f}^\lambda$, the empirical density of selection frequencies. Set $U=1$ (the U-shape indicator).\\
2. & Check U-shape.\\
   & 2.1. Check valley point.\\
   & \hspace{1cm} 2.1.1. Calculate $v_2=\operatorname{argmin}_x\tilde {f}^\lambda(x)$, the valley point position,\\
   & \hspace{2.1cm} where $\tilde {f}^\lambda $ is a smooth curve fitted based on $\hat{f}^\lambda$. (We use\\
   & \hspace{2.1cm} the R-function $\tt{smooth.spline ()}$, where the degree of \\
   & \hspace{2.1cm} freedom parameter is determined such that the derivative \\
   & \hspace{2.1cm} of $\tilde {f}^\lambda $ has only one sign change.)\\
   & \hspace{0.95cm} 2.1.2. \textbf{IF} $v_2>0.8$\\
        & \hspace{2.1cm} Set $U=0$, \textbf{GOTO} Step 3.\\
        & \hspace{2.1cm}\textbf{END IF}\\
   & 2.2. Calculate $v_1=\operatorname{argmax}_{x<v_2}\hat {f}^\lambda(x)$, the peak before $v_2$.\\
   & 2.3. Check if $\hat{f}^\lambda$ is ``roughly'' decreasing on $(v_1,v_2]$.\\
   & \hspace{1cm} 2.3.1. Calculate $\mu_1=(v_1+v_2)/2$, $s_1=\sum_{x\in [v_1,\mu_1]}\hat{f}^\lambda(x)$ and\\
   & \hspace{2.1cm} $s_2=\sum_{x\in [\mu_1,v_2]}\hat{f}^\lambda(x)$. \\
   & \hspace{0.95cm} 2.3.2. \textbf{IF} $s_1<s_2$\\
   & \hspace{2.1cm} Set $U=0$, \textbf{GOTO} Step 3.\\
   & \hspace{2.1cm}\textbf{END IF}\\
   & 2.4. Check if $\hat{f}^\lambda$ is ``roughly'' increasing on $(v_2,1]$.\\
   & \hspace{1cm} 2.4.1. Calculate $\mu_2=(v_2+1)/2$, $s_3=\sum_{x\in [v_2,\mu_2]}\hat{f}^\lambda(x)$ and\\
   & \hspace{2.1cm} $s_4=\sum_{x\in [\mu_2,1]}\hat{f}^\lambda(x)$. \\
   & \hspace{0.95cm} 2.4.2. \textbf{IF} $s_3>s_4$\\
   & \hspace{2.1cm} Set $U=0$, \textbf{GOTO} Step 3.\\
   & \hspace{2.1cm}\textbf{END IF}\\
3. & \textbf{RETURN} $v_1, v_2, U$.\\
\hline
\end{tabular}}\vspace*{6pt}

Therefore, if $\hat{f}^{\lambda}$ is not sufficiently large at the
tail, $\operatorname{FDR}\leq\alpha$ may not be achieved for a small value of $\alpha
$. The increasing trend also helps to obtain decent power since it
guarantees a substantial size of $S_{c}^{\lambda}$. Based on our
experience, the $\lambda$ values chosen based on (\ref{eq2.9}) and (\ref{eq2.10})
indeed always corresponds to a ``U-shaped'' empirical selection
frequency distribution.

Thus, we propose the following simple procedure for identifying
``U-shaped'' $\hat{f}^{\lambda}$'s to determine the proper
regularization range in practice. An illustration for this procedure is
given in Figure~\ref{fig3}.

\begin{remark}\label{rem2} Step 2.1 is based on our extensive simulation where we
find that a large value of $v_2$ often corresponds
to a too-small $\lambda$, yielding too many null edges with high
selection frequencies, which makes (\ref{eq2.12}) difficult to hold for
reasonably small FDR levels $\alpha$ (see Section D1 in the
supplemental article [Li et al. (\citeyear{Lietal})]). If $\hat{f}^{\lambda}$ is not
recognized as ``U-shaped'' for a large range of $\lambda$'s, we
would consider the data as lack of signals where a powerful
$S_{c}^{\lambda}$ is not attainable. One example is the empty
network (see Section~\ref{sec3.2} and Figure S-1 in the supplemental article [Li
et al. (\citeyear{Lietal})]).
\end{remark}

Sections~\ref{sec2.2}--\ref{sec2.4} provide a procedure for network inference based on
directly estimating FDR. We name the procedure as
\textit{BINCO}---Bootstrap Inference for
Network COnstruction, as we suggest to use bootstrap resamples. The
main steps are summarized below.

\subsection{Randomized lasso}\label{sec2.5}

For an $L_1$ regularized procedure $A(\lambda)$, the
\textit{proper condition}
(Section~\ref{sec2.3}) is
satisfied if $A(\lambda)$ is selection consistent, which usually
requires strong conditions, for instance, the well-known
\textit{irrepresentable condition}
under the lasso
regression setting [\citet{ZhaYu06}, \citet{Zou06}, \citet{YuaLin07}, \citet{Wai09}] or the so-called
\textit{neighborhood stability condition}
under the GGM setting [Meinshausen and
B\"{u}hlmann (\citeyear{MeiBuh06}), Peng et al. (\citeyear{PenZhoZhu09})].
Recently, Meinshausen and
B\"{u}hlmann (\citeyear{MeiBuh10}) proposed the
\textit{randomized lasso}, which is a procedure based on
randomly sampled regularization parameters. For example, the
randomized lasso version of
\textit{space}
would be
%
\begin{equation}\label{eq2.13}
\qquad L(Y,\theta,W)=\frac{1}{2}\Biggl\{\sum_{i=1}^p
\biggl\Vert Y_i-\sum_{j:j\not
=i}\sqrt{
\frac{\sigma_{jj}}{\sigma_{ii}}}\rho_{ij} Y_j\biggr\Vert^2\Biggr\}
+\lambda\sum_{1\leq i<j\leq p}|\rho_{ij}|/w_{ij},
\end{equation}
\textit{\begin{tabular}{l p{333pt}@{}}
\hline
\multicolumn{2}{c}{\textbf{BINCO procedure}} \\
\hline
1. & \textbf{INPUT} $\Lambda =(\lambda_1,\ldots,\lambda_k)$ the initial range of regularization parameter values; $Y_{n \times p}$ the
dataset; and $\alpha$ the desired FDR level. \\
2. & \textbf{FOR}  $i = 1$ \textbf{TO} $k$ \\
        & 2.1. $\lambda =\lambda _i$ \\
        & 2.2. Generate $\hat{f}^{\lambda }$ the empirical density of selection frequencies.\\
        & 2.3. Check whether $\hat{f}^{\lambda }$ is U-shaped based on the output $(v_1, v_2, U)$ from\\
        & \hspace{0.6cm} the ``U-Shape Detection Procedure.''\\
        & 2.4. \textbf{IF} $\hat{f}^\lambda $ is U-shaped (i.e., $U=1$) \\
        & \hspace{1cm} 2.4.1. Obtain the null density estimate  $\hat{f}^{\lambda }_0$ by (\ref{eq2.11}). \\
        & \hspace{0.95cm} 2.4.2. Find the optimal threshold $c^*(\lambda )$ by (\ref{eq2.9}), where the FDR \\
        & \hspace{1.9cm} is estimated based on (\ref{eq2.7}) with $f^{\lambda }$ and $f^{\lambda }_0$ replaced by $\hat{f}^{\lambda }$\\
        & \hspace{1.9cm} and $\hat{f}^{\lambda }_0$,respectively.\\
    & \hspace{1cm} 2.4.3. Obtain $S_{c^*(\lambda )}^\lambda $ and calculate $\hat{N_E}(S_{c^*(\lambda )}^\lambda)$, the estimated\\
        & \hspace{1.9cm} number of true edges  being selected, based on (\ref{eq2.8}).\\
        & \hspace{1cm}\textbf{END IF}\\
        & \hspace{1cm}\textbf{ELSE} $\hat{N_E}(S_{c^*(\lambda )}^\lambda)= 0$, $S_{c^*(\lambda              )}^\lambda=\emptyset$.\\
        & 2.5. \textbf{OUTPUT} $\hat{N_E}(S_{c^*(\lambda )}^\lambda)$ and $S_{c^*(\lambda                )}^\lambda$.\\
        & \textbf{NEXT} $i$\\
3. & Determine the optimal regularization $\lambda^*$ through (\ref{eq2.10}). The optimal selection is $S_{c^*(\lambda^* )}^{\lambda^*}$.\\
\hline
\end{tabular}}\vspace*{6pt}

\noindent where $w_{ij}$'s are randomly sampled from a probability
distribution $p(w)$ supported on $(l,1]$ for some $l\in(0,1]$
(note that $l=1$ corresponds to the ordinary%
 $L_1$ penalty). The
advantage of this randomized lasso procedure is that, by
perturbing
the regularization parameters, the irrelevant features may be
decorrelated from the true features in some configurations of
randomly sampled weights such that the irrepresentable condition is
satisfied. Therefore, it selects all true features with probability
tending to 1 and any irrelevant feature with a limiting probability
strictly less than 1. As a result, a consistent aggregation-based
procedure can be achieved under conditions ``typically much weaker
than the standard assumption of the irrepresentable condition''
[Meinshausen and B\"{u}hlmann (\citeyear{MeiBuh10}), Theorem~2]. For
this case, based on Lemma~\ref{le1}, the
\textit{proper condition}
is also satisfied.

If (\ref{eq2.13}) is used as the original (nonaggregated) procedure, an
additional parameter $l$, which controls the amount of perturbation
of the regularization parameter, needs to be chosen. A small $l$
guards better against false positives but damages power, while a
large $l$ may result in a liberal procedure. Here we provide a
two-step data-driven procedure for choosing an appropriate $l$ in
BINCO. We first fix $l=1$, that is, the ordinary $L_1$ penalty, to find
a proper range $\Lambda^*$ for $\lambda$ that corresponds to the
``U-shaped'' empirical mixtures. Then for each $\lambda\in\Lambda^*$, we
consider a set of pairs $\Lambda_2=\{(\lambda_i,l_i), i=1,\ldots,m\}$ such that
$\int_{l_i}^1\frac{\lambda_i}{w}p(w)\,dw=\lambda$, that is, keeping the
average amount of regularization unchanged. For example, in the
simulation study, we use $l_i=i/10, i=1,\ldots,9$. We then
pick the pair $(\lambda^*,l^*)\in\Lambda_2$ such that $l^*$ is
the smallest among those $l$'s that yield U-shaped empirical mixture
distributions. Our simulation shows that such a choice of $(\lambda^*,l^*)$ ensures good power for BINCO while controlling FDR in a
slightly conservative fashion.

\section{Simulation}\label{sec3}

In this section we first compare the performance of BINCO
with
\textit{stability selection}
[Meinshausen and B\"{u}hlmann (\citeyear{MeiBuh10})], and then
investigate the performance of BINCO with respect to various
factors, including the network structure, dimensionality, signal
strength and sample size.

We use
\textit{space}
[Peng et al. (\citeyear{PenZhoZhu09})] coupled with randomized lasso
(\ref{eq2.13}) as the
original nonaggregate procedure, where the random weights
$\frac{1}{w_{ij}}$'s are generated from the uniform distribution
$U[1,1/l]$ for $l\in(0,1]$. The selection frequencies are obtained
based on $B=100$ resamples. Since subsampling of size $[n/2]$ is
proposed for
\textit{stability selection}, we use subsampling
to generate resamples when comparing BINCO and
\textit{stability selection}. For investigating
BINCO's performance, we use bootstrap resamples because it
yields slightly better performance (see Remark~\ref{rem4}).

The performance of both methods are evaluated by true FDRs and power,
since for simulations we know whether an edge is true or null. In
addition, we define
\textit{ideal
power}, which is the best power one can
achieve for $S_c^{\lambda}$ given the true \mbox{$\operatorname{FDR}\leq \alpha$} (in
simulation we consider $\alpha=0.05$ and $\alpha=0.1$). Based on
\textit{ideal power}, we can evaluate the
efficiency of the methods under different settings. For each
simulation setting, results are based on 20 independent simulation
runs.

\subsection{Comparison between BINCO and stability selection}\label{sec3.1}

\textit{Stability selection}
procedure
selects $S_{\mathrm{stable}}^{\Lambda}(t)\equiv\{(i,j)\dvtx \operatorname{max}_{\lambda
\in\Lambda}(X_{ij}^{\lambda})\geq t\}$, a set of edges with the
maximum selection frequency over a prespecified regularization set
$\Lambda$ exceeding a threshold $t$. Assuming an exchangeability
condition upon the irrelevant variables (here the null edges),
Meinshausen and B\"{u}hlmann [(\citeyear{MeiBuh10}), Theorem~1] derived
an upper bound for the expected number of falsely selected variables
for each choice of $t>0.5$. Specifically, under suitable conditions,
the expected number of null edges selected by the set
$S_{\mathrm{stable}}^{\Lambda}(t)$, denoted by $E(V)$, satisfies
%
\begin{equation}\label{eq3.1}
E(V)\leq\frac{q_{\Lambda}^2}{(2t-1)N_{\Omega}},
\end{equation}
where $N_{\Omega}=p(p-1)/2$ is the total number of candidate edges and
$q_{\Lambda}$ is the expected number of edges selected under at least
one $\lambda\in\Lambda$. In practice, $q_{\Lambda}$ can be estimated
by $\frac{1}{B}\sum_{i=1}^B|\bigcup_{\lambda\in\Lambda}S^{\lambda
}(Y^{i})|$. Dividing both sides of (\ref{eq3.1})\vadjust{\goodbreak} by $|S_{\mathrm{stable}}^{\Lambda
}(t)|$, we obtain
%
\begin{equation}\label{eq3.2}
\frac{E(V)}{|S_{\mathrm{stable}}^{\Lambda}(t)|}\leq\frac{q_{\Lambda
}^2}{(2t-1)N_{\Omega}\cdot|S_{\mathrm{stable}}^{\Lambda}(t)|}.
\end{equation}
Although
\textit{stability selection}
is intended to control $E(V)$, for an easier comparison with BINCO, we
use $\frac{E(V)}{|S_{\mathrm{stable}}^{\Lambda}(t)|}$ to approximate FDR and
obtain the optimal $S_{\mathrm{stable}}^{\Lambda}(t)$ by finding the smallest
threshold $t$ such that the upper bound on the right-hand side of (\ref{eq3.2})
is less than or equal to $\alpha$.

For data generation, we first consider a
\textit{power-law
network}
with $p=500$ nodes whose degree
(i.e., the number of connected edges for each node) distribution
follows $P(k)\sim k^{-\gamma}$. The scaling exponent $\gamma$ is
set to be 2.3, which is consistent with the findings in the
literature for biological networks [\citet{New03}]. There are in
total 495 true edges in this network and its topology is illustrated
in Figure~\ref{fig32}(a). The sample size is $n=200$. Two settings with
different signal strengths are considered: (1) strong signal, the
mean and standard deviation (SD) of nonzero $|\rho_{ij}|$'s are
0.34 and 0.13, respectively; (2) weak signal, the mean and SD of
nonzero $|\rho_{ij}|$'s are 0.25 and 0.09, respectively. Note both
positive and negative correlations are allowed in this network.

We compare the performance of BINCO and
\textit{stability selection}
at a targeted FDR level of 0.05. For BINCO,
we consider $\Lambda_0=\{40,50,\ldots,100\}$ as the initial range
for $\lambda$ and then obtain the optimal final selection following
the steps at the end of Section~\ref{sec2.4}. For
\textit{stability
selection}, since no specific guidance was provided for
choosing $\Lambda$ and $l$ (the randomized lasso regularization
perturbation parameter), we consider three different values for
$l\in\{0.5,0.8,1\}$ and a collection of intervals
$\Lambda=(\lambda_{\operatorname{min}},\lambda_{\operatorname{max}})$ with $\lambda_{\operatorname{min}}$
varying from 40 to 100 and $\lambda_{\operatorname{max}}=100$. This choice
of $\Lambda$ is due to the fact that the upper bound in (\ref{eq3.2})
cannot be controlled at 0.05 for any $t$ for $\lambda_{\operatorname{min}}<40$, and the performance of
\textit{stability
selection}
is largely invariant for $\lambda_{\operatorname{max}}$.

\begin{figure}
\centering
\begin{tabular}{@{}c@{\hspace*{3pt}}c@{}}

\includegraphics{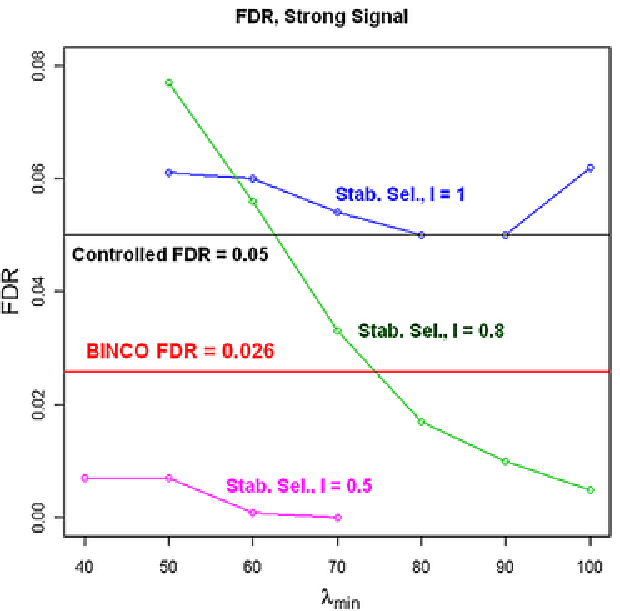}
 & \includegraphics{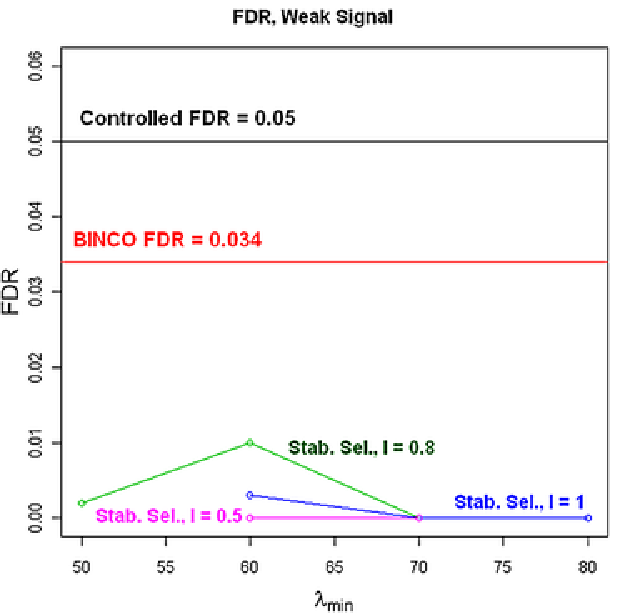}\\
\footnotesize{(a)} & \footnotesize{(b)}\\[6pt]

\includegraphics{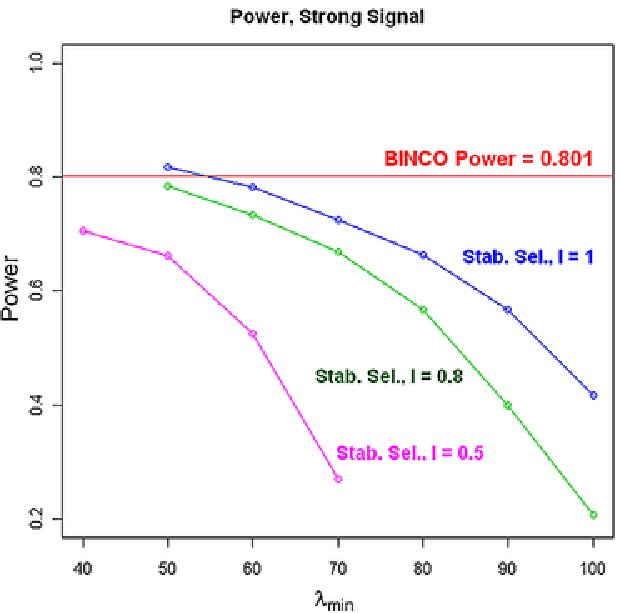}
 & \includegraphics{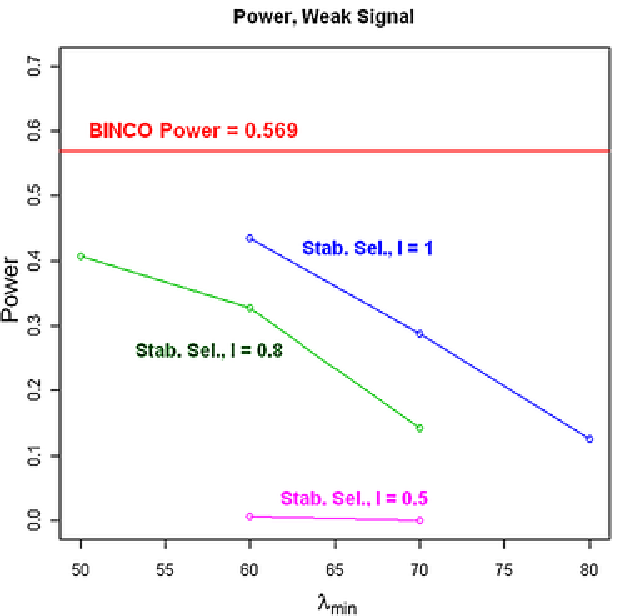}\\
\footnotesize{(c)} & \footnotesize{(d)}\vspace*{-2pt}
\end{tabular}
\caption{The FDR (top panels) and power (bottom panels) for BINCO and
stability selection
(Stab. Sel.). \textup{(a)} and \textup{(c)} are for the strong signal setting; \textup{(b)} and
\textup{(d)} are for the weak signal setting.} \label{fig31}\vspace*{-2pt}
\end{figure}

\begin{figure}
\centering
\begin{tabular}{@{}ccc@{}}

\includegraphics{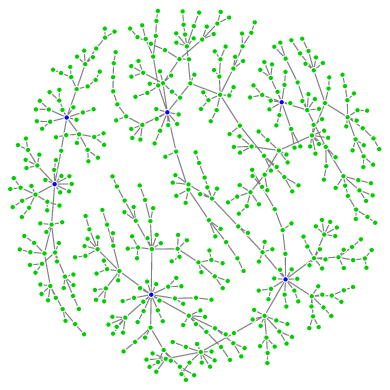}
 & \includegraphics{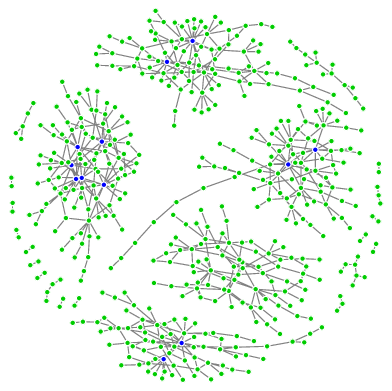} & \includegraphics{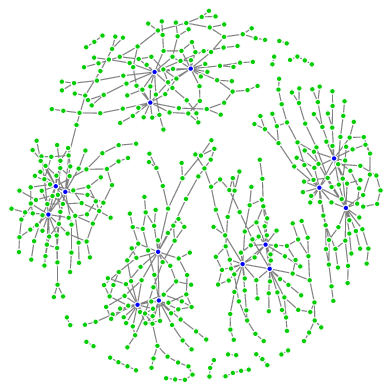}\\
\footnotesize{(a)} & \footnotesize{(b)} & \footnotesize{(c)}\vspace*{-2pt}
\end{tabular}
\caption{Different network topologies: \textup{(a)} Power-law network, number
of true edges${} = 495$; \textup{(b)}~Empirical network, number of true edges${}=633$;
\textup{(c)} Hub network, number of true edges${}=587$. All three networks have
$p=500$ nodes.} \label{fig32}\vspace*{-2pt}
\end{figure}

When the signals are strong, BINCO gives a conservative $\operatorname{FDR}=0.026$
but still maintains good power${}=0.801$ [Figures~\ref{fig31}(a) and~\ref{fig31}(c)]. The
performance of
\textit{stability selection}
varies for different choices of $\lambda_{\operatorname{min}}$ and $l$. The
FDRs are larger than the targeted level 0.05 for some $\lambda_{\operatorname{min}}$'s
when $l=0.8$ and for all $\lambda_{\operatorname{min}}$'s
when $l=1$. For other cases (some $\lambda_{\operatorname{min}}$'s when
$l=0.8$ and all $\lambda_{\operatorname{min}}$'s when $l=0.5$), the FDR
control is very conservative and the corresponding power is
consistently lower than BINCO. When the signals are weak,
\textit{stability selection}
is much more
conservative than BINCO and results in much lower power [Figures~\ref{fig31}(b)
and 4(d)]. In Table~\ref{table1} we report the
\textit{ideal
power}, the power for BINCO and the best power for
\textit{stability selection}
(among different
choices of $\lambda_{\operatorname{min}}$) under $l=0.5$, 0.8 and 1. We
also calculate the power efficiency as the ratio of the power for
the method over the
\textit{ideal power}, for
BINCO and
\textit{stability selection},
respectively. It can be seen that the power of BINCO\vadjust{\goodbreak} is close to the
\textit{ideal power}
for both levels of signal
strength, while
\textit{stability selection}
is
too conservative when the signal strength is weak. For more detailed
results, see Section A1 in the supplemental article [Li~et~al.
(\citeyear{Lietal})].\vspace*{-2pt}

\begin{remark}\label{rem3} In some cases we find that
\textit{stability selection}
fails to control FDR. We suspect this
may be due to the violation of the exchangeability assumption in
Theorem~1 of Meinshausen and B\"{u}hlmann (\citeyear{MeiBuh10}). We
examine the impact of the exchangeability assumption by simulation
and find that when it is violated,\vadjust{\goodbreak} the theoretical upper bound in
(\ref{eq3.1}) for $E(V)$ may not hold (see Section D2 in the supplemental
article [Li et al. (\citeyear{Lietal})] for further details).\vspace*{-2pt}
\end{remark}

\subsection{Further investigation of BINCO}\label{sec3.2}

Now we investigate the effects of the network structure,
dimensionality, signal strength and sample size on the performance of BINCO.
\begin{table}[b]\vspace*{-2pt}
\caption{Power comparison between BINCO and stability selection under
strong and weak signals}\label{table1}  
\begin{tabular*}{\textwidth}{@{\extracolsep{\fill}}lcccccc@{}}
\hline
&&&& \multicolumn{3}{c@{}}{\textbf{Stability selection}}\\[-4pt]
&&&& \multicolumn{3}{c@{}}{\hrulefill}\\
& &{\textbf{Ideal}\tnote{tt1}} & \textbf{BINCO} & $\bolds{l=1}$ & $\bolds{l=0.8}$ & $\bolds{l=0.5}$\\
\hline
{Strong signal}& Power & 0.853 & 0.801 & 0.818\tnote{tt3}
& 0.785\tnote{tt3} & 0.706 \\ 
& MPE\tnote{tt2} & 1\phantom{00.0} & 0.939 & 0.959\tnote{tt3} & 0.920\tnote{tt3}& 0.828
\\[3pt]
{Weak signal} &Power & 0.616 & 0.569 & 0.434\phantom{0} & 0.407\phantom{0} &
0.170 \\ 
& MPE\tnote{tt2} & 1\phantom{00.0} & 0.924 & 0.705\phantom{0} & 0.661\phantom{0} & 0.276 \\ 
\hline
\end{tabular*}
%
%
\tabnotetext[1]{tt1}{``Ideal'' refers to the \textit{ideal power} that can be achieved when the true distribution of null edges is known.}
\tabnotetext[2]{tt2}{Method Power Efficiency (MPE)${}={}$method power${}/{}$ideal power.}
\tabnotetext[3]{tt3}{FDR control failed.}
\end{table}

\textit{Network structure}.

We consider four different network topologies:
\textit{empty network},
\textit{power-law network},
\textit{empirical network}
and
\textit{hub network}. In each network there are five disconnected components with 100 nodes
each. Below is a brief description of the network topologies:

\begin{longlist}[(1)]
\item[(1)]
\textit{Empty network}: there is no edge connecting any pair of nodes.

\item[(2)]
\textit{Power-law network}: the degree follows a power-law distribution with parameter $\gamma
=2.3$ as described in Section~\ref{sec3.1}
[Figure~\ref{fig32}(a)].\vadjust{\goodbreak}

\item[(3)]
\textit{Empirical network}: the topology is simulated according to an empirical degree
distribution of one genetic regulatory network [\citet{Schetal05}] [Figure~\ref{fig32}(b)].

\item[(4)]
\textit{Hub network}: three nodes per component have a large number of connecting edges
($>$15) and all other nodes have a small number of connecting edges
($<$5) [Figure~\ref{fig32}(c)].
\end{longlist}

We set the sample size $n=200$. The signal strength for all networks
except for the empty network is fixed at the strong level as in
Section~\ref{sec3.1}.

\begin{table}
\tabcolsep=0pt
\caption{Investigation of the impact of different networks on BINCO
performance}\label{table2} 
\begin{tabular*}{\textwidth}{@{\extracolsep{\fill}}lcccccccc@{}}
\hline
&\multicolumn{4}{c}{\textbf{Targeted} $\bolds{\operatorname{FDR} = 0.05}$} & \multicolumn {4}{c}{\textbf{Targeted} $\bolds{\operatorname{FDR} =
0.10}$}\\[-4pt]
&\multicolumn{4}{c}{\hrulefill} & \multicolumn {4}{c@{}}{\hrulefill}\\
& \multicolumn{2}{c}{\textbf{FDR}} & \multicolumn{2}{c}{\textbf{Power}} &
\multicolumn{2}{c}{\textbf{FDR}} & \multicolumn{2}{c@{}}{\textbf{Power}} \\[-4pt]
& \multicolumn{2}{c}{\hrulefill} & \multicolumn{2}{c}{\hrulefill} &
\multicolumn{2}{c}{\hrulefill} & \multicolumn{2}{c@{}}{\hrulefill} \\
\textbf{Network topology}&\textbf{Mean}&\textbf{SD}&\textbf{Mean}&\textbf{SD}&\textbf{Mean}&\textbf{SD}&\textbf{Mean}&\textbf{SD}\\
\hline
Power-law & 0.046 & 0.009 & 0.810 & 0.013 & 0.096 & 0.013 & 0.845 &
0.013\\
Empirical & 0.032 & 0.019 & 0.523 & 0.040 & 0.068 & 0.034 & 0.565 &
0.040\\ 
Hub & 0.023 & 0.009 & 0.644 & 0.021 & 0.052 & 0.012 & 0.692 & 0.017\\
\hline
\end{tabular*}
\end{table}

\begin{table}[b]
\caption{Comparison of BINCO power and ideal power under different
networks}\label{table3} 
\begin{tabular*}{\textwidth}{@{\extracolsep{\fill}}lcccccc@{}}
\hline
&\multicolumn{3}{c}{\textbf{Targeted} $\bolds{\operatorname{FDR} = 0.05}$}& \multicolumn{3}{c@{}}{\textbf{Targeted}
$\bolds{\operatorname{FDR} = 0.10}$} \\[-4pt]
&\multicolumn{3}{c}{\hrulefill}& \multicolumn{3}{c@{}}{\hrulefill} \\
\textbf{Topology} & \textbf{Power-law} & \textbf{Empirical} & \textbf{Hub} & \textbf{Power-law} & \textbf{Empirical} & \textbf{Hub} \\
\hline
BINCO power & 0.810 & 0.523 & 0.644 & 0.845 & 0.565 & 0.692\\ %
Ideal power & 0.856 & 0.595 & 0.736 & 0.881 & 0.631 & 0.776\\
MPE\tnote{ttt1}& 0.946 & 0.879 & 0.875 & 0.959 & 0.895 & 0.892\\
[1ex] 
\hline
\end{tabular*}
\tabnotetext[1]{ttt1}{Method Power Efficiency (MPE)${}={}$method
power${}/{}$ideal power.}
\end{table}

For the empty network, the empirical mixture distributions of
selection frequencies monotonically decrease on a wide range of
$\lambda$ (Figure S-1) and are not recognized by BINCO as
``U-shaped.'' Thus, we reach the correct conclusion that there is no
signal in this case. In contrast, data sets from the other three
networks produce the desired ``U-shaped'' mixture distributions for
some $\lambda$ (Figure S-2).

We compare BINCO results across networks 2-4 with FDR targeted at
level $\alpha= 0.05$ and 0.1. BINCO gives slightly conservative
control on FDR and achieves reasonable power for all three networks
(Table~\ref{table2}). The comparison to the
\textit{ideal power}
shows that the network topologies investigated here have only a small
effect on BINCO's efficiency (Table~\ref{table3}).


%
\textit{Dimensionality}.
We investigate the impact of dimensionality on the performance of
BINCO. We consider the power-law network and let the number of nodes
$p$ vary from 500, 800 to 1000. To keep the complexity of each
component the same across different choices of $p$, we set the
component size constant, being 100, and the number of components $C
= p/100$. Again the sample size $n=200$ is used for all three cases
and the signal strength is fixed at the strong level as in Section~\ref{sec3.1}.

For all three choices of $p$, BINCO performs similarly (Table~\ref{table4}),
with slightly conservative FDR and power around 0.8. The
dimensionality does not demonstrate a significant impact on BINCO.
BINCO's result is also largely invariant when we compare networks of
differing numbers of components with $p$ fixed (such that component
size varies, see Section A3 in the supplemental article [Li et al.
(\citeyear{Lietal})]).

\begin{table}
\caption{Investigation of the impact of different dimensionality on
BINCO performance}\label{table4}
\begin{tabular*}{\textwidth}{@{\extracolsep{\fill}}lcccccccc@{}} 
\hline
&\multicolumn{4}{c}{\textbf{Targeted} $\bolds{\operatorname{FDR} = 0.05}$} & \multicolumn {4}{c}{\textbf{Targeted} $\bolds{\operatorname{FDR} =
0.10}$}\\[-4pt]
&\multicolumn{4}{c}{\hrulefill} & \multicolumn {4}{c@{}}{\hrulefill}\\
& \multicolumn{2}{c}{\textbf{FDR}} & \multicolumn{2}{c}{\textbf{Power}} &
\multicolumn{2}{c}{\textbf{FDR}} & \multicolumn{2}{c@{}}{\textbf{Power}} \\[-4pt]
& \multicolumn{2}{c}{\hrulefill} & \multicolumn{2}{c}{\hrulefill} &
\multicolumn{2}{c}{\hrulefill} & \multicolumn{2}{c@{}}{\hrulefill} \\
\textbf{Dimension} $\bolds{p}$&\textbf{Mean}&\textbf{SD}&\textbf{Mean}&\textbf{SD}&\textbf{Mean}&\textbf{SD}&\textbf{Mean}&\textbf{SD}\\
\hline
\phantom{0}500 & 0.046 & 0.009 & 0.810 & 0.013 & 0.096 & 0.013 & 0.845 & 0.013\\
\phantom{0}800 & 0.030 & 0.007 & 0.769 & 0.010 & 0.083 & 0.010 & 0.811 & 0.012\\
1000& 0.043 & 0.007 & 0.784 & 0.008 & 0.096 & 0.011 & 0.821 & 0.007\\
\hline
\end{tabular*} \vspace*{-2pt}
\end{table}
%

\begin{table}[b]\vspace*{-2pt}
\caption{Investigation of the impact of different signal strength on
BINCO performance}\label{tableS10} 
\begin{tabular*}{\textwidth}{@{\extracolsep{\fill}}lcccccccc@{}}
\hline
&\multicolumn{4}{c}{\textbf{Targeted} $\bolds{\operatorname{FDR} = 0.05}$} & \multicolumn {4}{c}{\textbf{Targeted} $\bolds{\operatorname{FDR} =
0.10}$}\\[-4pt]
&\multicolumn{4}{c}{\hrulefill} & \multicolumn {4}{c@{}}{\hrulefill}\\
& \multicolumn{2}{c}{\textbf{FDR}} & \multicolumn{2}{c}{\textbf{Power}} &
\multicolumn{2}{c}{\textbf{FDR}} & \multicolumn{2}{c@{}}{\textbf{Power}} \\[-4pt]
& \multicolumn{2}{c}{\hrulefill} & \multicolumn{2}{c}{\hrulefill} &
\multicolumn{2}{c}{\hrulefill} & \multicolumn{2}{c@{}}{\hrulefill} \\
\textbf{Signal strength}&\textbf{Mean}&\textbf{SD}&\textbf{Mean}&\textbf{SD}&\textbf{Mean}&\textbf{SD}&\textbf{Mean}&\textbf{SD}\\
\hline
Strong & 0.046 & 0.009 & 0.810 & 0.013 & 0.096 & 0.013 & 0.845 & 0.013\\
Weak & 0.032 & 0.010 & 0.579 & 0.024 & 0.063 & 0.014 & 0.617 & 0.018\\
Very weak & 0.035 & 0.026 & 0.252 & 0.040 & 0.065 & 0.037 & 0.310 &
0.039\\
\hline
\end{tabular*}
\end{table}

\textit{Signal strength}.
We consider three levels of signal strength: strong, weak and very
weak. The corresponding means and SDs of nonzero $|\rho_{ij}|$'s
are $(0.34, 0.13)$, $(0.25, 0.09)$ and $(0.21,0.07)$, respectively.
The network is the power-law network with $p=500$ and sample size is
$n=200$ for all settings.

BINCO provides good control on FDR, however, the power decreases from
0.8 to 0.3 as the signal weakens (Table~\ref{tableS10}). Comparing the power of
BINCO with the
\textit{ideal power} (Table~\ref{table33}), we
see that BINCO remains efficient and the loss in power is largely
due to reduction of signal strength.\vadjust{\goodbreak}

\begin{table}
\caption{Power comparison of BINCO power and ideal power when the
signal strength is strong, weak and very weak}\label{table33}
\begin{tabular*}{\textwidth}{@{\extracolsep{\fill}}lcccccc@{}}
\hline
&\multicolumn{3}{c}{\textbf{Targeted} $\bolds{\operatorname{FDR} = 0.05}$}& \multicolumn{3}{c@{}}{\textbf{Targeted}
$\bolds{\operatorname{FDR} = 0.10}$} \\[-4pt]
&\multicolumn{3}{c}{\hrulefill}& \multicolumn{3}{c@{}}{\hrulefill} \\
\textbf{Signal strength} & \textbf{Strong} & \textbf{Weak} & \textbf{Very weak} & \textbf{Strong} & \textbf{Weak} & \textbf{Very weak}
\\
\hline
BINCO power & 0.810 & 0.579 & 0.252 & 0.845 & 0.617 & 0.310\\ %
Ideal power & 0.856 & 0.615 & 0.279 & 0.881 & 0.651 & 0.345\\
MPE\tnote{tttt1}& 0.946 & 0.941 & 0.903 & 0.959 & 0.948 & 0.899\\
\hline
\end{tabular*}
\tabnotetext[1]{tttt1}{Method Power Efficiency (MPE)${} = {}$method power${} / {}$ideal
power.}\vspace*{-3pt}
\end{table}
%



%
\textit{Sample size}.
Now we consider the impact of sample size $n$ by varying it from
200, 500 and 1000, while keeping the signal strength at the ``very
weak'' level as in the previous simulation. The network structure is
again the power-law network with $p=500$.

With an increased sample size, the power of BINCO is significantly
improved from 0.3 to nearly 0.9 while the FDRs are well controlled
(Table~\ref{tableS11}).

\begin{table}[b]\vspace*{-3pt}
\caption{Investigation of the impact of different sample size on BINCO
performance}\label{tableS11}  
\begin{tabular*}{\textwidth}{@{\extracolsep{\fill}}lcccccccc@{}}
\hline
&\multicolumn{4}{c}{\textbf{Targeted} $\bolds{\operatorname{FDR} = 0.05}$} & \multicolumn {4}{c}{\textbf{Targeted} $\bolds{\operatorname{FDR} =
0.10}$}\\[-4pt]
&\multicolumn{4}{c}{\hrulefill} & \multicolumn {4}{c@{}}{\hrulefill}\\
& \multicolumn{2}{c}{\textbf{FDR}} & \multicolumn{2}{c}{\textbf{Power}} &
\multicolumn{2}{c}{\textbf{FDR}} & \multicolumn{2}{c@{}}{\textbf{Power}} \\[-4pt]
& \multicolumn{2}{c}{\hrulefill} & \multicolumn{2}{c}{\hrulefill} &
\multicolumn{2}{c}{\hrulefill} & \multicolumn{2}{c@{}}{\hrulefill} \\
\textbf{Sample size}&\textbf{Mean}&\textbf{SD}&\textbf{Mean}&\textbf{SD}&\textbf{Mean}&\textbf{SD}&\textbf{Mean}&\textbf{SD}\\
\hline
\phantom{0}200 & 0.035 & 0.026 & 0.252 & 0.040 & 0.065 & 0.037 & 0.310 & 0.039\\
\phantom{0}500 & 0.024 & 0.010 & 0.684 & 0.012 & 0.049 & 0.011 & 0.714 & 0.014\\
1000& 0.045 & 0.010 & 0.869 & 0.013 & 0.090 & 0.015 & 0.891 & 0.012\\
\hline
\end{tabular*}
\end{table}

In summary, BINCO has good control for FDR under a wide
range of scenarios. Its performance is shown to be robust for
networks with different topologies and dimensionalities, and its
efficiency is not influenced much even when the signal strength is
weak. As the sample size increases, the power of BINCO is improved
significantly.

\begin{remark}\label{rem4}We propose to use bootstrap over subsampling, as the
former appears to give slightly better power. Intuitively, bootstrap
contains more distinct samples [0.632$n$, \citet{Pat62}] than $[n/2]$
subsampling (0.5$n$). However, the difference we have observed is
rather small. For example, we compare the power over 20 independent
samples between bootstrap and $[n/2]$ subsampling under the power-law
network setting. For $\operatorname{FDR}=0.05$, the power is 0.810 for bootstrap and
0.801 for subsampling (compare Tables~\ref{table3} and~\ref{table1}); while for $\operatorname{FDR}=0.1$,
the power is 0.845 for bootstrap and 0.835 for subsampling [compare
Tables~\ref{table3} and~S-7 from Li et al.\vadjust{\goodbreak} (\citeyear{Lietal})]. This observation is in agreement with the
conclusions of several others [Menshausen and B\"{u}hlmann
(\citeyear{MeiBuh10}), \citet{Fre77}, B\"{u}hlmann and Yu (\citeyear{BuhYu02})].
\end{remark}

\section{A real data application}\label{sec4}

We apply the BINCO method to a microarray expression data set
of breast cancer (BC) [\citet{Loietal}] to build a gene expression network related to the disease. The
data
(\href{http://www.ncbi.nlm.nih.gov/geo/query/acc.cgi?acc=GSE6532}{http://www.ncbi.nlm.}
\href{http://www.ncbi.nlm.nih.gov/geo/query/acc.cgi?acc=GSE6532}{nih.gov/geo/query/acc.cgi?acc=GSE6532})
contains measurements of expression levels of 44,928 probes in tumor
tissue samples from 414 BC patients based on the Affymetrix Human
Genome U133A, U133B and U133 plus 2.0 Microarray platforms.

We preprocess the data as follows. First, a global normalization is
applied by centering the median of each array to zero and scaling
the
\textit{Median Absolute Deviation}
(MAD) to
one. Probes with standard deviation (SD) greater than the
$25\%$-trimmed mean of all SDs are selected. We further focus on a
subset of $1257$ probes for genes from cell cycle and DNA-repair
related \mbox{pathways}
(\href{http://peiwang.fhcrc.org/internal/papers/DNArepair\_CellCircle\_related.csv/view}{http://peiwang.fhcrc.org/internal/papers/DNArepair\_CellCircle\_}\break
\href{http://peiwang.fhcrc.org/internal/papers/DNArepair\_CellCircle\_related.csv/view}{related.csv/view}),
as these pathways have been shown to play significant
roles in BC tumor initiation and development. Clinical information
including age, tumor size, ER-status (positive or negative) and
treatment status (tamoxifen treated or not) is incorporated in the
analysis as ``fake genes'' since we are also interested in
investigating whether gene expressions are associated with these
clinical characteristics. Finally, we standardize each expression
level to have mean zero and SD one. The resulting data set has $p =
1261$ genes/probes (including four clinical variables) and $n=414$
tumor samples.

\begin{figure}

\includegraphics{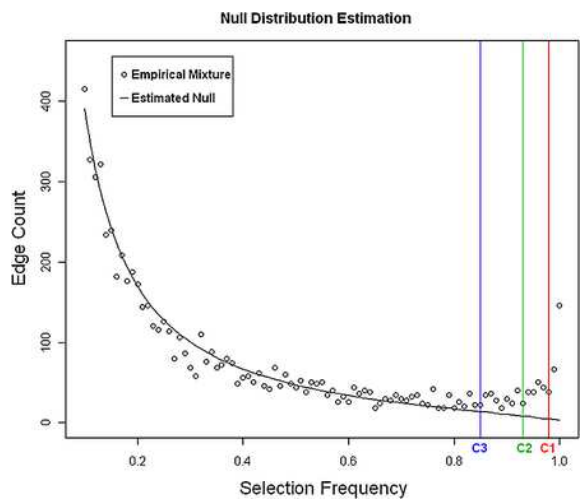}

\caption{The empirical selection frequency distribution of all edges
(dots) and the estimated selection frequency distribution of null edges
(solid line). The three vertical lines are drawn at the cutoffs
$C1=0.98$, $C2=0.93$ and $C3=0.85$ for FDR at 0.05, 0.1 and 0.2,
respectively.} \label{fig41}
\end{figure}

\begin{figure}

\includegraphics{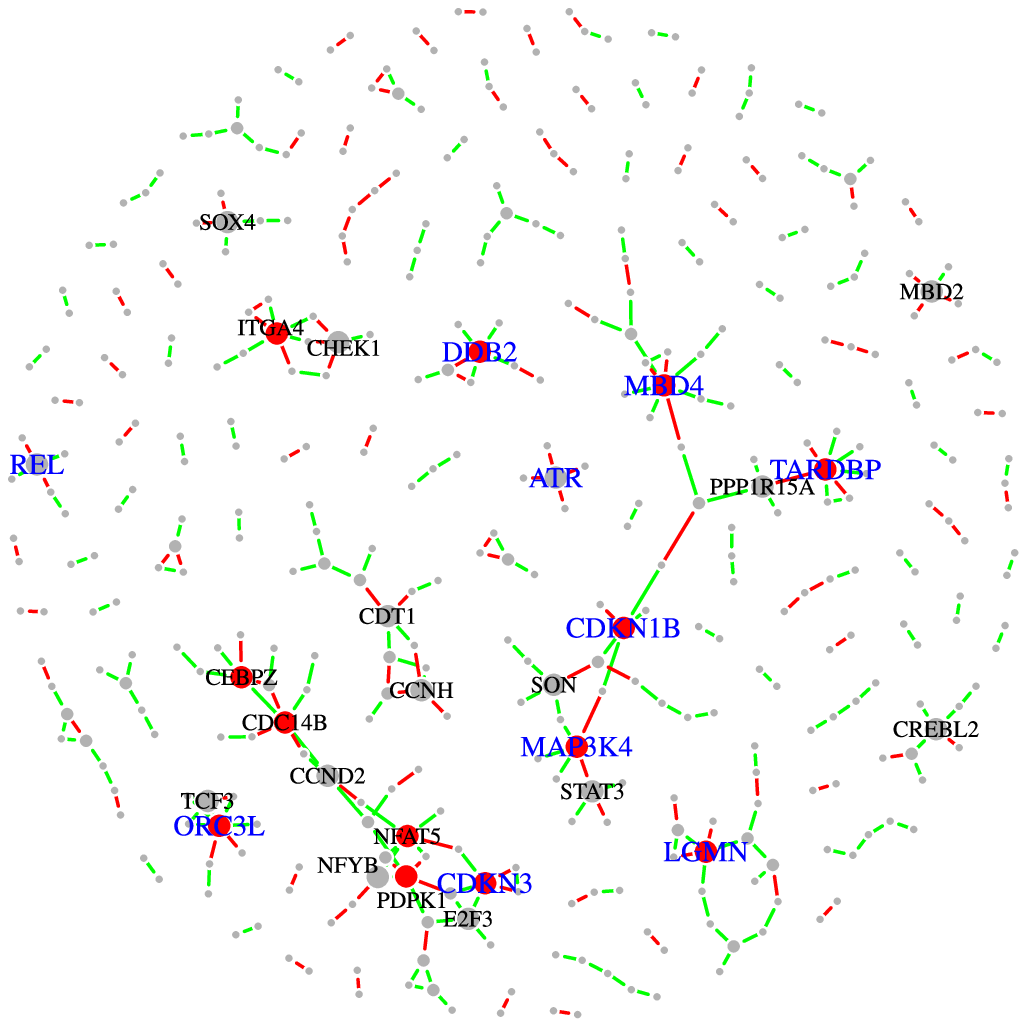}

\caption{Inferred networks at $\operatorname{FDR}=0.2$ from the BC expression data. A
total of 338 edges (selection frequencies $\geq0.85$) are identified.
Among these 338 edges, those with selection frequencies $\geq0.98$
(corresponding to the set with $\operatorname{FDR}=0.05$) are colored in red, while
other edges are colored in green. Genes with degree${}>$3 are labeled by
their symbols; genes with degree${}>$4 are indicated by red nodes. In
addition, the top ten genes with consistently high connection across
perturbed data sets are labeled in blue symbols.} \label{fig42}
\end{figure}

We generate selection frequencies by applying the
\textit{space}
algorithm with randomized lasso
regularization to $B=100$ bootstrap resamples. The initial range of
the tuning parameter $\lambda$ is set to be
$\Lambda=(100,120,\ldots,580)$. We then apply the BINCO procedure and
find that the optimal values for the regularization parameters are
$\lambda=340$ and $l=0.9$. The empirical distribution of selection
frequencies of all edges and the null density estimation are given
in Figure~\ref{fig41}. When the estimated FDR is controlled at 0.05, 0.1 and
0.2, BINCO identifies 125, 222 and 338 edges, respectively. The
estimated network for $\operatorname{FDR}=0.2$ is shown in Figure~\ref{fig42}. In this figure,
two components of a large connectivity structure are observed. They
contain most of the genes that are connected by a large number of
high-selection-frequency edges. This constructed network can help to
generate a useful biological hypothesis and to design follow-up
experiments to better understand the underlying mechanism in BC. For
example, BINCO suggests with high confidence for the association
between MAP3K4 and STAT3. MAP3K4 plays a role in the signal
transduction pathways of BC cell proliferation, survival and
apoptosis [\citet{BilJoh}], and the constitutive activation
of STAT3 is also frequently detected in BC tissues and cell lines
[Hsieh, Cheng and Lin (\citeyear{HsiCheLin05})]. Interestingly,\vadjust{\goodbreak}
both MAP3K4 and STAT3 play roles in the regulation of c-Jun, a novel
candidate oncogene whose aberrant expression contributes to the
progression of breast and other human cancers [Tront, Hoffman and Liebermann (\citeyear{TroHofLie06});
\citet{Shaetal11}]. The association between
MAP3K4 and STAT3 detected by BINCO suggests their potential
cooperative roles in BC. It is also worth noting that for the four
clinical variables, the only edge with high selection frequency is
the one between age and ER-status (selection frequency $=0.96$). All
edges between clinical variables and the genes/probes are
insignificant (selection frequencies $<0.12$).

Networks built on perturbed data sets can also be used to detect hub
genes (i.e., highly connected genes), which are often of great
interest due to the central role these genes may play in genetic
regulatory networks. The idea is to look for genes that show
consistent high connection in estimated networks across perturbed
data sets. Here, we propose to detect hub genes by the ranks of their
degrees based on the estimated networks using $\lambda=340$ and
$l=0.9$. The ten genes with the largest means and the smallest SDs
of the degree rank across 100 bootstrap resamples (see Figure S-3, in
black dots) are MBD4, TARDBP, DDB2, MAP3K4, ORC3L, CDKN1B, REL, ATR,
LGMN and CDKN3. Nine out of these ten genes have been reported
relevant to BC, while the remaining one (TARDBP) is newly discovered
to be related to cancer [Postel-Vinay et al. (\citeyear{Po12})], although its role in BC is not clear at
present. The neighborhood topologies of these hub genes in the
network estimated by BINCO are illustrated in Figure~\ref{fig42}. More details of
these hub genes are given in the supplemental article [Li et al. (\citeyear{Lietal})],
Section B.\looseness=-1\vadjust{\goodbreak}

\section{Discussion}\label{sec5}

In this paper we propose the BINCO procedure to conduct
high-dimensional network inference. BINCO employs model aggregation
strategies and selects edges by directly controlling the FDR. This
is achieved by modeling the selection frequencies of edges with a
two-component mixture model, where a flexible parametric
distribution is used to model the density for the null edges. By
doing this, BINCO is able to provide a good estimate of FDR and
hence properly controls the FDR. To ensure BINCO works, we propose a
set of screening rules to identify the U-shape characteristic of
empirical selection frequency distributions. Based on our
experience, a U-shape corresponds to a proper amount of
regularization such that the FDR is well controlled and the power is
reasonable. Extensive simulation results show that BINCO performs
well under a wide range of scenarios, indicating that it can be used
as a practical tool for network inference. Although we focus on the
GGM construction problem in this paper, BINCO is applicable to a
wide range of problems where model selection is needed because it
provides a general approach to modeling the selection frequencies.

We use a mixture distribution with two components, one corresponding
to true edges and the other corresponding to null edges, to model
the selection frequency distribution. This two-component mixture
model is adequate as long as the distribution of the null component
is identifiable and can be reasonably estimated, as formalized in
the
\textit{proper condition}. Note that the
\textit{proper condition}
holds for a wide range
of commonly used (nonaggregation) selection procedures (Lemma~\ref{le1},
Remark~\ref{rem1}). To further ensure the FDR can be controlled at a
reasonable level, we propose a U-shape detection procedure and only
apply BINCO if the empirical distribution of selection frequencies
passes the detection. These rules for U-shape detection are
empirical but appear to work very well based on our extensive
simulations.

BINCO works well despite the presence of correlations between edges
(see Section D1 in the supplemental article [Li et al. (\citeyear{Lietal})]), because we
use the independence of edges only as a working assumption . It is
well known that if the marginal distribution is correctly specified,
the parameter estimates are consistent even in the presence of
correlation. This is similar to the generalized estimating
equations, where if the mean function is correctly specified, the
parameters will be consistently estimated [Liang and Zeger (\citeyear{LZ86})].
Toward this end, we use the three-parameter power beta distribution
to allow for adequate flexibility in modeling the marginal
distribution of selection frequencies.

BINCO is computationally feasible for high-dimensional data. The
major computational cost lies in generating the selection
frequencies via resampling. For each resample, the computational
cost is determined by that of the nonaggregated procedure BINCO
coupled with. In terms of
\textit{space}, it is
$O(np^2)$. The processing time for a data set with $n=200$, $p=500$,
under a given $\lambda$ and 100 bootstrap samples to generate
selection frequencies is about 20 minutes on a PC with Pentium
dual-core CPU at 2.8~GHz and 1~G ram. These selection frequencies can
be simultaneously generated through parallel computing for different
$\lambda$'s and weights. Fitting the mixture model takes much less
time, which is about 2 minutes for the above example on the same
computer.

Although we use GGM as our motivating example, BINCO works well even
if the multivariate normality assumption does not hold. Note that
the multivariate normality assumption only concerns the
interpretation of the edges. Under GGM, the presence of an edge
means conditional dependency of the corresponding nodes given all
other nodes. Without the normality assumption, one can only conclude
nonzero partial correlation between the two nodes given the rest of the
nodes. The
\textit{space} method used in this paper is to
estimate the concentration network (where an edge is drawn between
two nodes if the corresponding partial correlation is nonzero) and
has been shown to work well under nonnormal cases such as
multivariate-$t$ distributions [Peng et al. (\citeyear{PenZhoZhu09})]. We also generate data from nonnormal
distributions and found that BINCO works well in this situation (see
Section D4 in the supplemental article [Li et al. (\citeyear{Lietal})]).

BINCO is an aggregation-based procedure. In principle, it can be
coupled with any selection procedure. In this sense, it has a wide
range of applications as long as the features are defined (e.g.,
edges as in this paper, variables or canonical correlations as in
the example below) and the selection procedure is reasonably good,
for example, producing probabilities that satisfy the condition in
Lemma~\ref{le1}.
One application beyond GGM could be on the multi-attribute network
construction where the links/edges are defined based on canonical
correlations [Waaijenborg, Verselewel de Witt Hamer and Zwinderman (\citeyear{Wa08}),
Katenka and Kolaczyk (\citeyear{KatKol}),
Witten, Tibshirani and Hastie (\citeyear{WitTibHas09})]. Another interesting
extension may be on the time-varying network construction [\citet{Koletal10}] where appropriate
incorporation of the time-domain structure across aggregated models
will be important. These are beyond the scope of this paper and will
be pursued in future research.

The R package BINCO is available through CRAN.

\begin{appendix}
\section*{\texorpdfstring{Appendix: Proof of Lemma~\lowercase{\protect\ref{le1}}}{Appendix: Proof of Lemma 1}}\label{app}
\begin{pf}
Suppose as the sample size $n$ increases, an edge selection procedure
$A(\lambda)$ gives selection probabilities $\{\tilde{p}^{(n)}_{ij}\}$
(with respect to resample space) which uniformly satisfy
\renewcommand{\theequation}{\Alph{section}.\arabic{equation}}
\setcounter{equation}{0}
\begin{equation}\label{equA1}
\tilde{p}^{(n)}_{ij}\rightarrow1\qquad  \mbox{if }(i,j)\in E
\end{equation}
and
%
\begin{equation}\label{equA2}
\limsup\tilde{p}^{(n)}_{ij}\leq M<1\qquad \mbox{if }(i,j)\in
E^c.
\end{equation}

Suppose $B$ is large such that $\frac{B+1}{B}M<1$. Let $X$ be a random
sample from the set of selection frequencies $\{X_{ij}^{\lambda}\}$
generated by applying $A(\lambda)$ on $B$ resamples, that is,
$\operatorname{Pr}(X=X_{ij}^{\lambda})=1/N_{\Omega}$, $(i,j)\in\Omega$. Also
suppose $X$ has density $f_{ij}^{\lambda}$ if $X=X_{ij}^{\lambda}$.
Then the mixture model (\ref{eq2.6}) becomes
%
\begin{eqnarray}
 f^{\lambda}(x)&=& (1-\pi)f_0^{\lambda}(x)+\pi
f_1^{\lambda}(x)
\nonumber
\\[-8pt]
\\[-8pt]
\nonumber
&=&\sum_{(i,j)\in E^c}
\frac{1}{N_{\Omega}}f_{ij}^{\lambda}(x)+\sum
_{(i,j)\in
E}\frac{1}{N_{\Omega}}f_{ij}^{\lambda}(x)
\end{eqnarray}
with $(1-\pi)f_0^{\lambda}(x)=\sum_{(i,j)\in E^c}\frac{1}{N_{\Omega
}}f_{ij}^{\lambda}(x)$ and $\pi f_1^{\lambda}(x)=\sum_{(i,j)\in
E}\frac{1}{N_{\Omega}}f_{ij}^{\lambda}(x)$.\vadjust{\goodbreak}

Because of the i.i.d. nature of resamples given the data,
$f_{ij}^{\lambda}$ is a binomial density with $\tilde{p}_{ij}^{(n)}$
as the probability of success, that is, $f_{ij}^{\lambda}(x) ={B
\choose k} (\tilde{p}_{ij}^{(n)})^k(1-\tilde{p}_{ij}^{(n)})^{B-k}$
for $x=k/B,k=0,1,\ldots,B$. This binomial density is monotone
decreasing for $x$ greater than its mode $\mu_{ij}=\frac{[(B+1)\tilde
{p}_{ij}^{(n)}]}{B}$ or $\frac{[(B+1)\tilde{p}_{ij}^{(n) }]-1}{B}$. By
(\ref{equA2}), given $V_1=\frac{B+1}{B}M <1$ and $\varepsilon>0$ such that
$V_1+\varepsilon<1$, $\exists N$ such that for all $n>N$ $\operatorname
{max}_{(i,j)\in E^c}(\mu_{ij})<V_1+\varepsilon$ and hence for any null
edge $(i,j)\in E^c$, $f_{ij}^{\lambda}(x)$ is monotone decreasing on
$[V_1+\varepsilon,1]$, which implies C2 since $f_0^{\lambda}(x)=\frac
{1}{(1-\pi)N_{\Omega}}\sum_{(i,j)\in E^c}f_{ij}^{\lambda}(x)$.
Also,\vspace*{2pt}
(\ref{equA1}) implies, for $(i,j)\in E$, $f_{ij}^{\lambda}(x)\rightarrow0$
uniformly for $x<1$, which implies C1 for any $V_2<1$. Taking $V_2$
such that $V_1<V_2<1$ satisfies the
\textit{proper condition}
and completes the proof.
\end{pf}
\end{appendix}

\section*{Acknowledgments}
We thank anonymous reviewers and editors for helpful comments that
significantly improved this paper. We also thank Ms. Noelle Noble
for technical editing.

\begin{supplement}[id=suppA]
\stitle{Supplement to ``Bootstrap inference for network construction
with an \mbox{application} to a breast cancer microarray study''\\}
\slink[doi]{10.1214/12-AOAS589SUPP}  
\sdatatype{.pdf}
\sfilename{aoas589\_supp.pdf}
\sdescription{This supplement contains additional
simulation results, details of the hub genes detected by BINCO on
the breast cancer data, and examples of $p_{ij}$ and
$\tilde{p}_{ij}$ being close.}
\end{supplement}


%

\printaddresses

\end{document}